\documentclass[pra,twocolumn,showpacs,amsmath,amssymb,floatfix]{revtex4}
\usepackage{graphicx}
\usepackage{dcolumn}
\usepackage{bm}

\begin{document}

\title{On the qubit limit of cavity nonlinear optics}
\author{Hideo Mabuchi}
\affiliation{Edward L.\ Ginzton Laboratory, Stanford University, Stanford, California 94305, USA}

\date{December 30, 2011}

\begin{abstract}
Many proposals for solid-state photonic implementations of quantum information processing utilize high-quality optical resonators to achieve strong coupling between guided fields and heterogeneously incorporated qubits. Given the practical difficulty of accurately placing quantum dots, vacancy centers, or other such atom-like emitters throughout a complex nanophotonic circuit, it would be natural to consider whether high-quality resonators could be used in conjunction with bulk optical nonlinearities to create optically-coupled qubit degrees of freedom via lithographic patterning of a homogeneous medium. A recent limit theorem for quantum stochastic differential equations can be used to prove rigorously that this should be possible, in principle, using resonators incorporating a strongly Kerr-nonlinear, $\chi^{(2)}$-nonlinear, or two-photon absorbing material with very low loss at the fundamental optical wavelength.
\end{abstract}

\maketitle

\noindent Optical resonators incorporating nonlinear refractive or absorptive materials have long been of interest in classical information technology for photonic computation and switching~\cite{Gibb85}. Recent advances in the fabrication of high-quality nanophotonic resonators via lithographic patterning of optically nonlinear materials hold great promise for such applications, enabling the study of optical bistability and related switching phenomena at very low (sub-femtojoule) energy scales~\cite{Noza10,Kuma10}. In the attojoule (few-photon) limit that one might ultimately hope to reach with strongly nonlinear low-loss materials and ultra-high quality resonator fabrication, cavity nonlinear optics in fact crosses over into a regime of nonclassical physics characterized by quantum fluctuations and squeezing~\cite{Carm08}. Should we therefore expect some kind of quantum dynamics to take the place of classical (dissipative) switching in this limit? In the case of an intracavity Kerr nonlinearity it can be shown that the two stable branches of dispersive optical bistability limit to the two basis states of a coherent qubit that responds canonically to optical fields driving the resonator. A qubit limit can also in principle be reached with $\chi^{(2)}$ nonlinearity and, somewhat surprisingly, with two-photon absorption as long as low losses are maintained at the fundamental wavelength of the cavity resonance. It is hoped that this observation may help further to motivate research on lithography-compatible materials for nonlinear optics.

Our proof of the above assertions relies on a recent limit theorem for quantum stochastic differential equations (QSDE's) due to Bouten, Van Handel and Silberfarb~\cite{Bout08}. We therefore begin by establishing notation required to connect with the technical setup of this theorem, however, the reader need not be familiar with QSDE's to follow the main results below. The cavity nonlinear optical models that we wish to consider can be specified via choice of coefficients in a Hudson-Parthasarathy equation~\cite{Huds84,Barc06} for the time evolution of a Heisenberg propagator. The coefficients, conventionally labeled $K$, $L_i$ and $N_{ij}$, are generally operator-valued with $i$ and $j$ as indices for the $n$ input-output channels coupled to the cavity. The Hamiltonian of the intracavity dynamics is $H={\rm Im}\,[K]$ and the corresponding master equation for the intracavity dynamics is given by $(\hbar=1)$
\begin{equation}
\dot{\rho}=-i[H,\rho]+\sum_{i=1}^n\left\{ L_i^\dag\rho L_i - \frac{1}{2}L_iL_i^\dag\rho - \frac{1}{2}\rho L_iL_i^\dag\right\}.
\end{equation}
Note that the $L_i$ as defined here and in~\cite{Bout08} are hermitian conjugates of the usual Lindblad operators from quantum optics. Although the $N_{ij}$ do not appear in the master equation for the internal degrees of freedom, they are required to construct scattering models~\cite{Goug09a,Goug09b} in which the cavity input-output channels drive or are driven by additional quantum systems~\cite{Carm93,Gard93}.

Within this framework, we first consider a single-mode optical resonator incorporating a Kerr nonlinear medium. The corresponding Hamiltonian can be written~\cite{Wall08}
\begin{equation}
H=\Delta\,a^\dag a + \chi\,a^\dag a^\dag aa=(\Delta-\chi+\chi a^\dag a)a^\dag a,
\label{eq:Kerr2}
\end{equation}
where $a$ is the annihilation operator for the intracavity field, $\Delta$ is the detuning of the cavity resonance from the frequency $\omega_p$ of a rotating frame for the model, and $\chi$ is a coefficient that fixes the strength of the Kerr nonlinearity. The second form of the Hamiltonian emphasizes that a Kerr nonlinearity induces a shift in the cavity resonance frequency proportional to the intracavity photon number. If we consider a cavity that has two input-output channels ({\it i.e.}, is coupled to two waveguides, one of which could be used to model distributed losses) with associated photon coupling rates $\kappa_{a1,2}$, we can choose $N_{ij}=\delta_{ij}$ and
\begin{equation}
L_1=e^{i\omega_p t}\sqrt{\kappa_{a1}}a^\dag,\quad L_2=e^{i\omega_p t}\sqrt{\kappa_{a2}}a^\dag.
\end{equation}
As can be seen by assembling the corresponding master equation, the total cavity decay rate is thus set to $\kappa_a\equiv\kappa_{a1}+\kappa_{a2}$. If we want to consider the dynamics with a coherent driving field incident through channel $1$, we can use the series product~\cite{Goug09a,Goug09b} to derive the simple modifications
\begin{eqnarray}
H&\rightarrow&\Delta\,a^\dag a + \chi\,a^\dag a^\dag aa + \frac{\sqrt{\kappa_{a1}}}{2i}
(\alpha a^\dag - \alpha^* a),\nonumber\\
L_1&\rightarrow&e^{i\omega_p t}(\sqrt{\kappa_{a1}}a^\dag + \alpha^*),\;L_2\rightarrow e^{i\omega_p t}\sqrt{\kappa_{a2}}a^\dag.\label{eq:Kerrmodel}
\end{eqnarray}
Here $\alpha$ is the complex amplitude of the driving field ($\vert\alpha\vert^2$ has units of photons/sec) and we assume that the rotating frame of the model has the same frequency as the driving field. With this notation in place we can remind the reader of the physical origin of dispersive optical bistability in the classical regime, in which $\kappa_a\gg\vert\chi\vert$. Assuming $\Delta$ and $\chi$ have opposite sign and that $\vert\Delta\vert\agt\kappa_a$, there can be two equilibrium conditions for fixed $\kappa_{a1,2},\alpha,\Delta,\chi$: either $\langle a^\dag a\rangle$ is small because the cavity is driven off-resonance and the detuning `correction' $\chi a^\dag a$ is negligible, or $\langle a^\dag a\rangle$ becomes large enough to make $(\Delta-\chi+\chi a^\dag a) \alt\kappa_a$ and the external driving field can self-consistently maintain the high intracavity photon number because of the decrease in effective detuning.

\begin{figure}[tb]
\includegraphics[width=0.45\textwidth]{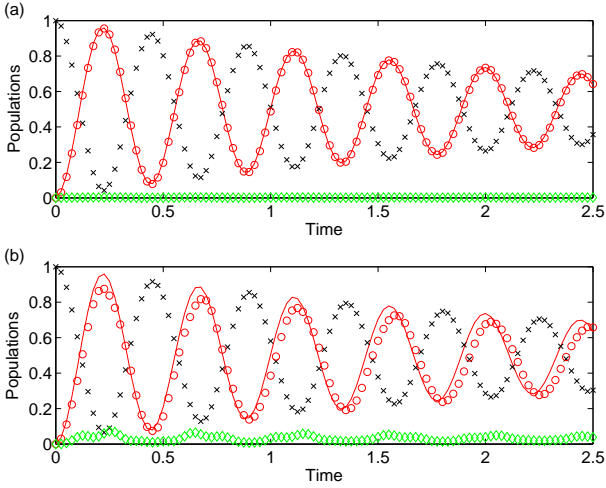}
\caption{\label{fig:Kerreg} Numerical integrations of the pre-limit (finite $\chi$) master equation for a driven Kerr-nonlinear cavity with $\kappa_{a1}=0.5$, $\kappa_{a2}=\Delta=0$, $\alpha=10$ and (a) $\chi=-100$, (b) $\chi=-20$. Black crosses are $\langle 0\vert\rho\vert 0\rangle$, red circles are $\langle 1\vert\rho\vert 1\rangle$, green diamonds are the sum of the remaining populations, red solid curve is $\langle 1\vert\rho\vert 1\rangle$ according to the limit (qubit) master equation~(\ref{eq:qubitME}).}
\vspace{-0.1in}
\end{figure}

If we now contemplate the possibility of achieving $\vert\chi\vert\gg\kappa_a$ in future nanophotonic systems with strongly nonlinear media and high-quality resonators, we immediately recognize that by setting $\Delta=\chi$ in Eq.~(\ref{eq:Kerr2}) we would be left with a model of a highly anharmonic oscillator. If the level spacing between the one- and two-photon basis states is thus made very different from that of the spacing between the zero- and one-photon states, it would seem to follow that a finite driving field should drive transitions only within the lowest `qubit' subspace~\cite{Imam97}. This type of scenario recalls some basic ideas from circuit quantum electrodynamics~\cite{Mart02} in which the nonlinear inductance of a Josephson Junction can be used to isolate a qubit subspace of a superconducting LC-oscillator in a similar fashion.

\begin{figure}[tb]
\includegraphics[width=0.45\textwidth]{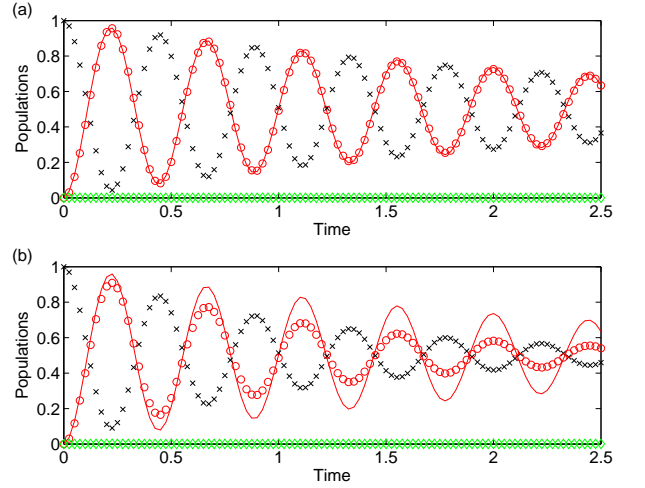}
\caption{Numerical integrations of the pre-limit (finite $g$, $\kappa_b$) master equation for a driven $\chi^{(2)}$-nonlinear cavity with $\kappa_{a1}=0.5$, $\kappa_{a2}=\Delta=0$, $\kappa_{b}=5000$, $\alpha=10$ and (a) $g=-3000$, (b) $g=-600$. Black crosses are $\langle 0_a 0_b\vert\rho\vert 0_a 0_b\rangle$, red circles are $\langle 1_a 0_b\vert\rho\vert 1_a 0_b\rangle$, green diamonds are the sum of the remaining populations, red solid curve is $\langle 1\vert\rho\vert 1\rangle$ according to the limit (qubit) master equation~(\ref{eq:qubitME}).\label{fig:chi2eg}}\vspace{-0.1in}
\end{figure}

To develop this idea more rigorously, we invoke Theorem 11 from~\cite{Bout08} to obtain the limit QSDE that governs the open quantum system dynamics in the parameter regime $\vert\chi\vert\gg\kappa_a,\alpha,\Delta$. We satisfy the `singular scaling assumption' and `structural requirements' for the theorem by setting (in the notation of~\cite{Bout08})
\begin{eqnarray}
Y&=&i\chi a^\dag a^\dag aa,\quad A=0,\quad F_1=F_2=0,\nonumber\\
{\cal H}_0&=&{\rm span}\,\{\vert 0\rangle,\vert 1\rangle\},\nonumber\\
{\tilde Y}&=&\sum_{m=2}^\infty\frac{-i}{m(m-1)\chi}\vert m\rangle\langle m\vert,\label{eq:Kerrsetup}
\end{eqnarray}
where $\vert m\rangle$ denotes the $m$-photon Fock state. We thus obtain a limit master equation that can be written
\begin{eqnarray}
\dot{\rho}&=&-i[H',\rho] + \kappa_a\left\{\sigma\rho\sigma^\dag - \frac{1}{2}\sigma^\dag\sigma\rho
- \frac{1}{2}\rho\sigma^\dag\sigma\right\},\nonumber\\
H'&=&\Delta\sigma^\dag\sigma - i\sqrt{\kappa_{a1}}(\alpha\sigma^\dag-\alpha^*\sigma),\label{eq:qubitME}
\end{eqnarray}
where $\sigma\equiv\vert 0\rangle\langle 1\vert$ and the evolution of the cavity state is now confined to the qubit subspace, ${\rm span}\,\{\vert 0\rangle,\vert 1\rangle\}$. We clearly recognize this master equation as being equivalent to that of a two-level atom with spontaneous emission rate $\kappa_a$, driven by an external field with Rabi frequency $2\sqrt{\kappa_{a1}}\vert\alpha\vert$ and detuning $\Delta$. Hence the emergence of qubit-like dynamics is established in this limit. Fig.~\ref{fig:Kerreg} shows numerical examples~\cite{Tan99} of the behavior of the pre-limit (finite $\chi$) master equation with $\kappa_{a1}=0.5$, $\kappa_{a2}=\Delta=0$, $\alpha=10$. With $\chi=-100$ we observe a damped Rabi oscillation within the qubit subspace in close correspondence with quantitative predictions of the limit model, and for $\chi=-20$ there is still qualitative agreement although some population leaks quickly out of the qubit subspace.

A qubit limit can also be derived in the case of an intracavity $\chi^{(2)}$-type nonlinearity~\cite{Carm08} that degenerately couples the high-quality cavity mode of interest to a much lower-quality pump mode at twice the frequency~\cite{Male05}. Starting from the pre-limit coefficients $N_{ij}=\delta_{ij}$ and
\begin{eqnarray}
L_1&=&e^{i\omega_p t}(\sqrt{\kappa_{a1}}a^\dag + \alpha^*),\nonumber\\
L_2&=&e^{i\omega_p t}\sqrt{\kappa_{a2}}a^\dag,\quad L_3=e^{2i\omega_p t}\sqrt{\kappa_b}b^\dag,\nonumber\\
H&=&\Delta a^\dag a + \frac{\sqrt{\kappa_{a1}}}{2i}(\alpha a^\dag - \alpha^* a)
+2\Delta b^\dag b\nonumber\\
&&+ i\frac{g}{2}(a^\dag a^\dag b - aab^\dag),\label{eq:chi2ME}
\end{eqnarray}
where $b$ is now the annihilation operator for a pump mode at twice the frequency of the fundamental, we first consider the limit of large $g$ and $\kappa_b$ with $g^2/\kappa_b$ fixed. Note that we assume no driving of the pump cavity mode and assign it only a single input-output channel for decay. Applying once again the theorem from~\cite{Bout08} with
\begin{eqnarray}
Y&=&-\frac{1}{2}\kappa_b b^\dag b,\quad A=\frac{g}{2}(aab^\dag - a^\dag a^\dag b),\nonumber\\
F_1&=&F_2=0,\quad F_3=e^{2i\omega_p t}\sqrt{\kappa_b}b^\dag,\nonumber\\
{\cal H}_0&=&{\rm span}\,\{\vert 0_a 0_b\rangle,\vert 1_a 0_b\rangle,\ldots\},\nonumber\\
{\tilde Y}&=&I_a\otimes\sum_{m_b=1}^\infty\frac{-2}{m_b\kappa_b}\vert m_b\rangle\langle m_b\vert,\label{eq:Chi2setup}
\end{eqnarray}
where $I_a$ denotes the identity operator on the Fock space of the $a$ mode, we obtain the limit master equation
\begin{eqnarray}
\dot{\rho}&=&-i[H',\rho]+\kappa_a\left\{a\rho a^\dag - \frac{1}{2}a^\dag a\rho - \frac{1}{2}\rho a^\dag a
\right\}\nonumber\\
&&+\frac{g^2}{\kappa_b}\left\{aa\rho a^\dag a^\dag - \frac{1}{2}a^\dag a^\dag aa\rho - \frac{1}{2}
\rho a^\dag a^\dag aa\right\},\nonumber\\
H'&=&\Delta a^\dag a - i\sqrt{\kappa_{a1}}(\alpha a^\dag - \alpha^* a),\label{eq:tpaME}
\end{eqnarray}
with the $b$ mode now adiabatically eliminated~\footnote{Model~(\ref{eq:Kerrmodel}) with $\chi\rightarrow -g^2/4\Delta_b$ can be obtained instead by replacing $2\Delta b^\dag b$ with $\Delta_b b^\dag b$ in~(\ref{eq:chi2ME}) and $g,\Delta_b\rightarrow\infty$.}. We can obtain a final master equation identical to Eq.~(\ref{eq:qubitME}) by using the theorem again to take the limit of large $g^2/\kappa_b$, setting
\begin{eqnarray}
Y&=&-\frac{g^2}{2\kappa_b}a^\dag a^\dag aa,\quad A=0,\nonumber\\
F_1&=&F_2=0,\quad F_3=e^{2i\omega_p t}\frac{g}{\sqrt{\kappa_b}}a^\dag a^\dag,\nonumber\\
{\cal H}_0&=&{\rm span}\,\{\vert 0_a 0_b\rangle,\vert 1_a 0_b\rangle\},\nonumber\\
{\tilde Y}&=&I_b\otimes\sum_{m_a=2}^\infty\frac{-2\kappa_b}{m_a(m_a-1)g^2}\vert m_a\rangle\langle m_a\vert,\label{eq:TPAsetup}
\end{eqnarray}
where $I_b$ is the identity operator of the $b$ mode.

\begin{figure}[tb]
\includegraphics[width=0.45\textwidth]{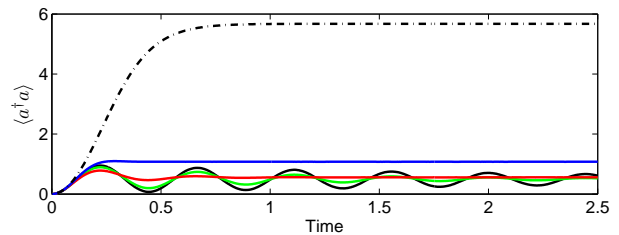}
\caption{(a) Time evolution of $\langle a^\dag a\rangle$ for the TPA master equation~(\ref{eq:tpaME}) with $\kappa_{a1}=0.5$, $\kappa_{a2}=0$, $\Delta=0$, $\alpha=10$ and $g^2/\kappa_b=5000$ (black), $200$ (green), $40$ (red), $8$ (blue), $0.5$ (black dash-dot).\label{fig:tpaeg2}}\vspace{-0.1in}
\end{figure}

In view of the latter result, we see that we could actually {\em start} with a model~(\ref{eq:tpaME}) with (cavity enhanced~\cite{Lin10,delV11}) two-photon absorption (TPA) as the only optical nonlinearity and still expect to obtain qubit-like dynamics in the limit of large TPA rate. This is encouraging from a practical perspective since nonlinear absorption seems generally more straightforward to engineer~\cite{Jaco09,Saha11,Yang11} than nonlinear refraction without concomitant TPA~\cite{Chri10}. The idea of using strong TPA to confine the quantum evolution of the intracavity field state to the zero/one-photon subspace is reminiscent of several recent proposals for quantum logic gates~\cite{Fran07,Huan11} or all-optical switching~\cite{Jaco09,Huan10} based on the `quantum Zeno' effect. Fig.~\ref{fig:chi2eg} shows numerical examples generated by integrating the pre-limit $\chi^{(2)}$ master equation corresponding to the model~(\ref{eq:chi2ME}); very similar results are obtained by integrating the pre-limit TPA master equation~(\ref{eq:tpaME}) with corresponding values of $g^2/\kappa_b$ (not shown). In both cases the apparent decay rate of the Rabi oscillation increases when the relevant parameter ($g$ or $g^2/\kappa_b$) is not very large, although the confinement to the qubit subspace remains quite good; the manner of approach to the qubit limit is thus seen to be somewhat different in the $\chi^{(2)}$/TPA scenarios than for the Kerr nonlinearity. Fig.~\ref{fig:tpaeg2} displays plots of $\langle a^\dag a\rangle$ as a function of time, generated by numerical integration of the TPA master equation~(\ref{eq:tpaME}) for $g^2/\kappa_b$ decreasing from $5000$ down to $0.5$, illustrating the loss of Rabi oscillations and a transition to monotonic evolution qualitatively similar to that of a linear cavity. It is interesting to note however that with $g^2/\kappa_b\rightarrow 0$ the steady-state photon number with $\kappa_a=0.5$ and $\alpha=10$ would be $\langle a^\dag a\rangle\approx 800$, hence the optical limiting effect of TPA on the response of the fundamental mode to a coherent external drive is seen to be remarkably strong even for the lowest TPA rate considered $(g^2/\kappa_b=0.5=\kappa_a)$. Indeed, relatively weak (in the context of the current discussion) TPA in bulk semiconductors is widely investigated in classical photonics for the design of broadband optical limiters~\cite{VanS88} and related logic devices for pulsed signal fields~\cite{Haya11}.

The natural information processing paradigm of purely linear photonics is that of Laplace transfer functions acting on classical analog signals~\cite{Mell02}, while the addition of moderate optical nonlinearities enables bistable switching phenomena that can be used to implement classical digital logic~\cite{Gibb85}. Here we have seen that the incorporation of extreme optical nonlinearities (while preserving low loss at the fundamental wavelength) could in principle enable a transition to quantum information processing~\cite{Niel00} with localized photonic qubits~\cite{Mabu11a}. It seems natural to ask however if there should be useful `quantum analog' signal representations that would naturally suit the regime of strong but finite optical nonlinearity, and whether a corresponding `semi-quantum' information processing paradigm could be of practical interest as an intermediate stage in the evolution of nanophotonic technologies.

\begin{figure}[tb]
\includegraphics[width=0.45\textwidth]{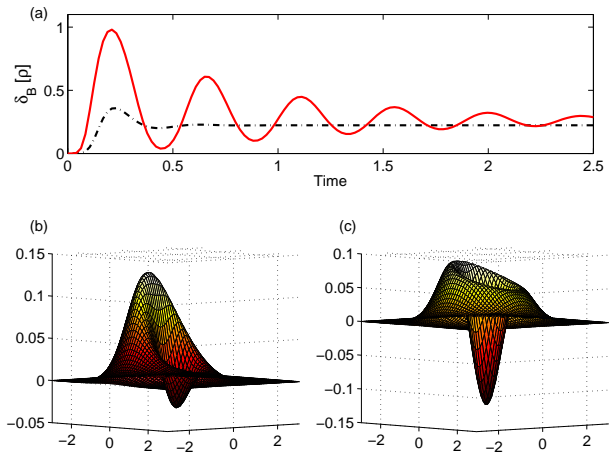}
\caption{(a) Time evolution of $\delta_B[\rho]$, a relative-entropy based non-Gaussianity measure~\cite{Geno10,Geno08}, for the pre-limit TPA model with $\kappa_{a1}=0.5$, $\kappa_{a2}=0$, $\Delta=0$ and $g^2/\kappa_b=200$ (red solid trace) or $g^2/\kappa_b=20$ (black dash-dot trace). (b) Wigner function of the intracavity field at $t=0.2101$ for $g^2/\kappa_b=20$. (c) Wigner function of the intracavity field for $g^2/\kappa_b=200$.
\label{fig:nG}}\vspace{-0.1in}
\end{figure}

While a serious exploration of the latter idea is beyond our current scope, it may be worth noting that strong but finite cavity-enhanced TPA could be a useful resource for generating non-Gaussian states in nanophotonic circuits, which potentially could be used to implement ideas from continuous-variable quantum information theory~\cite{Geno10}. Fig.~\ref{fig:nG}(a) displays $\delta_B[\rho]$, a relative entropy-based measure~\cite{Geno10,Geno08} of the non-Gaussianity of the intracavity field state as a function of time, computed by numerical integration of the pre-limit TPA master equation~(\ref{eq:tpaME}) with $\kappa_{a1}=0.5$, $\kappa_{a2}=\Delta=0$ and $g^2/\kappa_b=200$ (red solid trace) or $g^2/\kappa_b=20$ (black dash-dot trace), starting from $\vert 0\rangle$ at $t=0$. The Wigner functions of the intracavity field states near the point of maximum non-Gaussianity around $t\approx 0.21$ are shown in Figs.~\ref{fig:nG}(b,c). It is interesting to note that with $g^2/\kappa_b=20$, for which the intracavity state of Fig.~\ref{fig:nG}(b) has nearly 10\% population in $\vert 2\rangle$, the dissipative nonlinearity of TPA can still induce a negative component of the Wigner function even in a regime of imperfect confinement to the qubit subspace.

HM thanks M.~Fejer for discussions on nonlinear optics, and the 2011 Les Houches session on Quantum Machines for inspiration. This work is supported by DARPA/N66001-11-1-4106 and NSF/PHY-1005386.

\end{document}